\documentclass[11pt,amsmath,amssymb,prd,onecolumn]{revtex4-1}
\usepackage{isotope}
\usepackage{color}
\usepackage{graphicx}
\usepackage{slashed}

\newcommand{\be}{\begin{equation}}
\newcommand{\ee}{\end{equation}}
\newcommand{\eq}[1]{Eq.\,\eqref{eq:#1}}

\newcommand{\fig}[1]{Fig.\,\ref{fig:#1}}
\newcommand{\tab}[1]{Table\,\ref{tab:#1}}
\newcommand{\tabs}[2]{Tables\,\ref{tab:#1} and \ref{tab:#2}}
\newcommand{\ssec}[1]{Sec.\,\ref{sec:#1}}

\begin{document}

\title{Photon and neutron production as in-situ diagnostics of proton-boron fusion}

\author{B. M. Hegelich}
\email{Authors alphabetical}
\affiliation{Department of Physics, The University of Texas, Austin, 78712, USA}
\author{L. Labun}
\email[Corresponding author email: ]{labun@utexas.edu}
\affiliation{Department of Physics, The University of Texas, Austin, 78712, USA}
\author{O. Z. Labun}
\affiliation{Department of Physics, The University of Texas, Austin, 78712, USA}
\author{T. A. Mehlhorn}
\affiliation{Mehlhorn Engineering Consulting Services, Beaverton OR USA}
\affiliation{HB11 Energy Holdings Pty, 11 Wyndora Ave, Freshwater, NSW 2096, AUSTRALIA}
\date{10 November 2022}

\begin{abstract}
Short-pulse, ultra high-intensity lasers have opened new regimes for studying fusion plasmas and creating novel ultra-short ion beams and neutron sources.  Diagnosing the plasma in these experiments is important for optimizing the fusion yield but difficult due to the picosecond time scales, 10s of micron-cubed volumes and high densities.  We propose to use the yields of photons and neutrons produced by parallel reactions involving the same reactants to diagnose the plasma conditions and predict the yields of specific reactions of interest.  In this work, we focus on verifying the yield of the high-interest aneutronic proton-boron fusion reaction $\isotope[11]{B}(p,2\alpha)\isotope[4]{He}$, which is difficult to measure directly due to the short stopping range of the produced $\alpha$s in most materials.  We identify promising photon-producing reactions for this purpose and compute the ratios of the photon yield to the $\alpha$ yield as a function of plasma parameters.  In beam fusion experiments, the $\isotope[11]{C}$ yield is an easily-measurable observable to verify the $\alpha$ yield.  In light of our results, improving and extending measurements of the cross sections for these parallel reactions are important steps to gaining greater control over these laser-driven fusion plasmas.
\end{abstract}

\maketitle

\section{Introduction}



Short-pulse lasers offer new experimental approaches to creating and studying fusion plasmas.  In contrast to long-pulse lasers which have been a primary tool in inertial-confinement fusion (ICF), short-pulse lasers have pulse durations $<1$ ps and use small focal spots to obtain peak intensities up to $10^{23}$ W/cm$^2$ in a single pulse.  Short-pulse lasers deliver their energy to the plasma in a time much shorter than the typical expansion timescale, and both electrons and ions achieve much higher momenta.  These plasma conditions are far from the quasi-thermal equilibrium of ICF, where burn has recently been achieved \cite{zylstra2022burning}, and the question is open whether or not the dynamics admit a pathway to net energy gain \cite{Mehlhorn2022}.  Short pulse lasers have successfully driven high-yield beam fusion experiments \cite{labaune2013fusion,baccou2015new}, which can in turn be translated into novel high-flux, ultra short-pulse ion \cite{giuffrida2020high,margarone2020generation,bonvalet2021energetic} and neutron sources \cite{norreys1998neutron,bang2013temperature,pomerantz2014ultrashort,jiao2017tabletop}.  

Short-pulse lasers can drive fusion in two ways: direct irradiation of a target containing the fusion reactants or laser-ion acceleration creating an ion beam that is dumped into a catcher/target.  To our knowledge, no experiment can claim to have optimized the fusion yield, and the efficiencies of these two methods for different candidate fusion reactions remains a topic of research.  Naively, one expects direct irradiation to convert laser energy more efficiently into fusion yield, in part because fusion can occur both in the neighbourhood of the focus where all ion species are heated, and in the colder bulk of the target by ions accelerated out of the focal region.  Anecdotally, recent experiments support this hypothesis \cite{Mehlhorn2022}.  

Proper optimization will require greatly improved understanding and control of experimental outcomes compared to current capabilities.  However the same laser properties, namely ultra short-pulse and typically small $\sim (10\mu\mathrm{m})^3$ focal volume, make the plasma difficult to diagnose.   Most interpretation is based on inference from the measured particle yields and spectra, sometimes supported by Monte Carlo or numerically-expensive kinetic laser-plasma simulations.  Improving experimental diagnostics of laser-driven nuclear reactions has thus become a significant topic of discussion \cite{infn2022}.  Our goal in this work is to identify new diagnostics providing information on the plasma conditions and nuclear reaction dynamics.

Out of the reactions studied with short-pulse lasers, we focus on the proton-boron-11 fusion reaction $\isotope[11]{B}(p,2\alpha)\isotope[4]{He}$, which is of particular interest because it releases $\simeq 8$ MeV into kinetic energy of the $3$ $\alpha$ particles and no neutrons.  The $\alpha$ particles themselves provide the most direct measure of the fusion yield, but because they deposit their kinetic energy into the surrounding medium very efficiently, only a small fraction of those produced escape the target.  This problem is especially acute in direct irradiation experiments \cite{belyaev2004neutron,belyaev2005observation, margarone2014advanced,picciotto2014boron, giuffrida2020high, kong2022alpha, margarone2022target}, where the mean kinetic energy and density of the medium vary by orders of magnitude in different regions of the target, precluding a systematic analytic correction for $\alpha$ stopping.

Since direct diagnostics of plasma conditions, such as probe laser pulses or atomic spectroscopy, remain an enormous technical challenge, we investigate other nuclear reactions for products whose yield or spectrum can be measured more reliably.  Photons and neutrons are the best candidates, but not all reactions will yield enough photons or neutrons that can be identified as coming from a specific reaction.  We introduce the yield ratio as a phenomenological tool to relate an easily-measured yield to the yield of interest.  Yield ratios have an established history in diagnosing the ICF plasma, where they can determine several of the important $\rho R$ parameters \cite{frenje2020nuclear}. We have previously used the ratio between $\alpha$ and $\isotope[11]{C}$ yields to determine the $\isotope[11]{B}(p,2\alpha)\isotope[4]{He}$ reaction yield more accurately in beam-target experiments \cite{McCary2022}, and here we demonstrate its utility in direct irradiation experiments as well.  The yield ratio eliminates normalization unknowns such as the local density of reactants, effective reaction volume and time, and takes as input a few model parameters, such as the mean ion kinetic energy, that can be determined from particle diagnostics.  We conclude by identifying the two best candidate reactions for proxy measurements of the $\isotope[11]{B}(p,2\alpha)\isotope[4]{He}$ yield in direct irradiation experiments and confirm $\isotope[11]{C}$ as the best proxy in beam-target experiments.  


\section{Accessible reactions}

The goal is to predict the outcomes and analyse the data from experiments on the aneutronic proton-boron fusion reaction $\isotope[11]{B}(p,2\alpha)\alpha$.  The $\isotope[11]{B}(p,2\alpha)\alpha$ cross section reaches $\sim 1$\,b around 650 keV center-of-mass (CM) energy, significantly higher than DD or DT fusion reactions because of the higher charge of boron.  In fact, most other proton-boron reactions require even higher CM energy before the cross section approaches 100mb, and the high cross section of $\isotope[11]{B}(p,2\alpha)\alpha$ in the $E<1$ MeV range is due to two identified resonances, related to above-threshold excited states of $\isotope[12]{C}$ \cite{AJZENBERGSELOVE19901}.  Recent work has resolved apparent normalization discrepancies in the measured cross section \cite{sikora2016new}, resulting in re-evaluation of the process as a candidate for fusion energy.  

With its cross section reaching $\sim 1$\,b already at 650 keV CM energy, lower than the thresholds of many other proton-initiated reactions on boron, $\isotope[11]{B}(p,2\alpha)\alpha$ is expected to have the highest yield in the laser-driven beam-fusion experiments.  
With higher laser intensities $I\gtrsim 10^{20}$ W/cm$^2$ though, the proton beam in the experiment can provide energies up to $\sim 50$ MeV \cite{McCary2022}, allowing many additional reactions that are naturally grouped as ``primary'' or ``secondary''.  Primary reactions are initiated by the protons scattering on $\isotope[11]{B}$, $\isotope[10]{B}$ or $\isotope[14]{N}$ as present in typical boron or boron-nitride solid targets.  These are listed in \tab{primaryreactions} with peak cross section, the corresponding CM energy at which the peak cross section is found and the range of CM energy over which data is available.  Secondary reactions are re-scattering of the $\alpha$ particles on the boron and nitrogen nuclei most prevalent in the environment.  These are listed in \tab{secondaryreactions}, similarly to the primary reactions.  The databases contain additional processes, such as $\isotope[14]{N}(p,n+p)\isotope[13]{N}$ , $\isotope[14]{N}(p,n+\isotope[3]{He})\isotope[11]{C}$, but the data is too sparse and the larger number of fragments generally makes the $Q$ values for such reactions negative and large in magnitude.  Consequently, their cross sections should have somewhat higher thresholds suppressing their contributions to the yields.

\begin{table}
\begin{tabular}{l|c|c|c|c}
\hline
Reaction  & $Q$ [MeV] & $\sigma_{\rm max}$[b] & $E_{cm}^{\rm (max\sigma)}$[MeV] & Data range\\\hline
$\isotope[11]{B}(p,2\alpha)\isotope[4]{He}$ & 8.59,5.65,8.68 & 0.8 & 0.6 & $0.03 < E_{cm} < 19$\,MeV \\
$\isotope[11]{B}(p,\alpha)\isotope[8]{Be}$ & $8.1^*$ & 0.01 & 1.0 & $0.03 < E_{cm} < 1$\,MeV\\
$\isotope[11]{B}(p,n)\isotope[11]{C}$ & $-2.765$ & 0.4 & 7  & $2.5 < E_{cm} < 15$\,MeV  \\
$\isotope[11]{B}(p,\gamma)\isotope[12]{C}$ & 15.9  & $3\times 10^{-6}$ & 14 & $7 < E_{cm} < 25$\,MeV \\
$\isotope[10]{B}(p,\alpha)\isotope[7]{Be}$ & 1.147 & 0.5 & 5.0 & $0.07 < E_{cm} < 11.15$\,MeV \\
$\isotope[10]{B}(p,n)\isotope[10]{C}$ & $-4.94^*$ & 0.004$^\mathbf{a}$ & 8 & $4 < E_{cm} < 16$\,MeV \\
$\isotope[10]{B}(p,\gamma)\isotope[11]{C}$ & 8.69 & 0.0003 & 6 & $1.8 < E_{cm} < 6$\,MeV \\
$\isotope[14]{N}(p,\alpha)\isotope[11]{C}$ & $-2.92$ & 0.29 & 7 & $3.78 < E_{cm} < 22$\,MeV \\
$\isotope[14]{N}(p,n)\isotope[14]{O}$ & $-6.44^*$ & 0.01$^\mathbf{b}$ & 9.5 & $6.3 < E_{cm} < 30$\,MeV\\ \hline
\end{tabular}
\caption{\label{tab:primaryreactions} Primary reactions in the range of proton energies.  The range of CoM energy for which data is available as well as the maximum cross section and its CoM energy are given for numerical comparison.  All data retried from sources in the EXFOR and JANIS databases, with the exception of $\isotope[11]{B}(p,2\alpha)\isotope[4]{He}$ for which we use the concordant normalization of Ref. \cite{sikora2016new}.  $Q$ values marked with an $^*$ are computed from the mass difference of the initial and final states.  Others are from the literature.  $^\mathbf{a}$ A recent measurement \cite{ALVES2000899} finds $\sim 5\times$ larger cross section than several previous measurements.  To be conservative, we quote the older but more consistent results here.  $^\mathbf{b}$ There appear two distinct groups of cross sections measurements: two experiments find the cross section roughly 10$\times$ larger than the majority of others.}
\end{table}

\begin{table}
\begin{tabular}{l|c|c|c|c}
Reaction & $Q$ [MeV] & $\sigma_{\rm max}$[b] & $E_{cm}^{\rm (max\sigma)}$[MeV]  & Data available\\\hline
$\isotope[11]{B}(\alpha,p)\isotope[14]{C}$ & $1.3^*$ & 0.04 & 3.6 & $0.6 < E_{cm} < 36$\,MeV\\
$\isotope[11]{B}(\alpha,n)\isotope[14]{N}$ & $0.16^*$ & $4\times 10^{-4}$ & 1.5 & $0.4 < E_{cm} < 1.76$\,MeV \\
$\isotope[10]{B}(\alpha,p)\isotope[13]{C}$ & 3.85 & -- & -- & No data \\
$\isotope[10]{B}(\alpha,n)\isotope[13]{N}$ & 1,\,3.51 & 0.1 & 4.6 & $2.55 < E_{cm} < 4.7$\,MeV\\
$\isotope[14]{N}(\alpha,n)\isotope[17]{F}$ & $-4.74^*$ & 0.12 & 9.2 & $6.3 < E_{cm} < 20$\,MeV\\
$\isotope[14]{N}(\alpha,\gamma)\isotope[18]{F}$ & 4.42 & -- & -- &  No data \\ \hline
\end{tabular}
\caption{\label{tab:secondaryreactions} Secondary reactions in the range of $\alpha$ particle energies.  The range of CoM energy for which data is available as well as the maximum cross section and its CoM energy are given for numerical comparison.  $Q$ values marked with an $^*$ are computed from the mass difference of the initial and final states.  Others are from the literature. }
\end{table}

\begin{figure}
\includegraphics[width=0.48\textwidth]{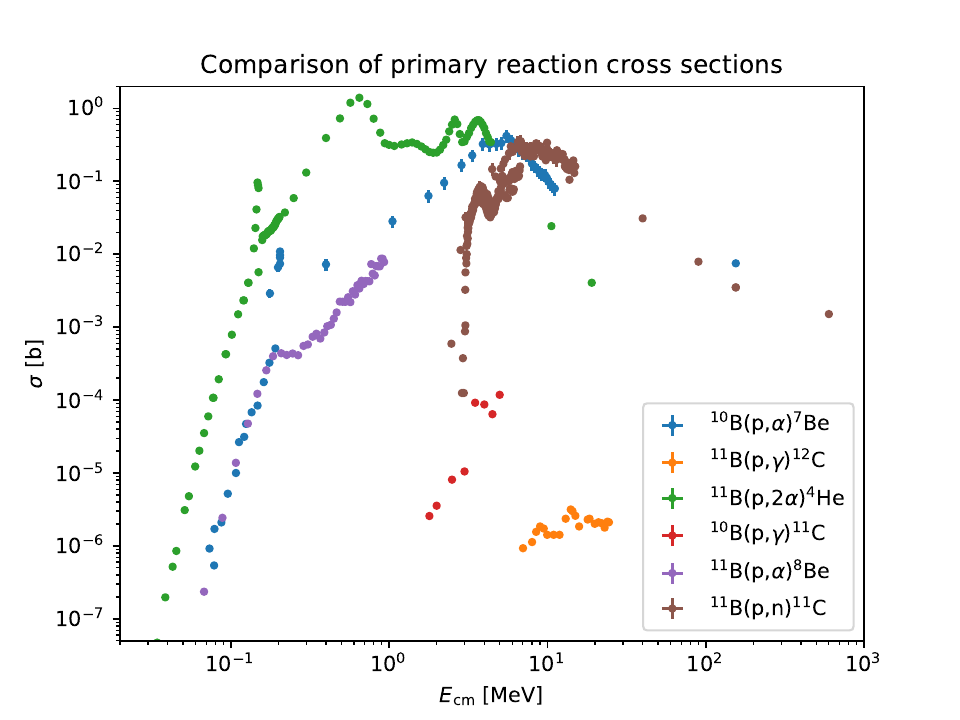}
\includegraphics[width=0.48\textwidth]{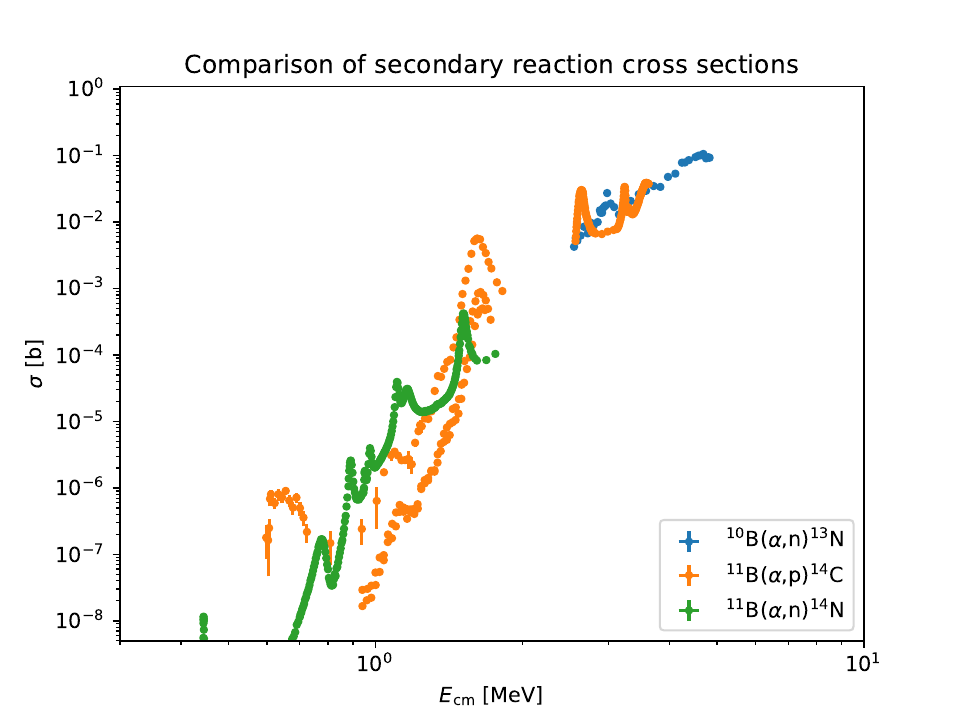}
\caption{\label{fig:sigmacomp} Cross sections of primary reactions (left) and secondary reactions (right).  Data from EXFOR with all sources combined.  Note the scales are logarithmic, and the extent of the energy range is different on each.}
\end{figure}

The corresponding cross sections are plotted in Figure \ref{fig:sigmacomp}.  The global sets from EXFOR \cite{zerkin2018experimental} frequently include inconsistent measurements, as for example the recently resolved normalization in the $\isotope[11]{B}(p,2\alpha)\isotope[4]{He}$ cross section \cite{sikora2016new}.  Cross section data is one significant source of uncertainty in our yield predictions.  We have plotted the global data sets for each cross section without distinguishing their sources and in our calculations, we will use fits to these global data with a few exceptions described in the captions of \tabs{primaryreactions}{secondaryreactions}.  We do not attempt to model the cross sections outside the range of available data.  Instead, for numerical integrations we implement best fit curves that are forced rapidly to zero outside the range of experimental data.  This choice almost certainly underestimates the yield for several processes.   Notably the cross sections of $\isotope[11]{B}(p,\alpha)\isotope[8]{Be}$, $\isotope[10]{B}(p,\gamma)\isotope[11]{C}$ and all secondary reactions for which data is available seem likely to continue to increase with CM energy.  However, the lack of information precludes quantifying the uncertainty in any attempted modeling of the cross section.  

Among these primary and secondary reactions, we identify promising candidates to be diagnostics.  As mentioned above, the difficulty in verifying the $\isotope[11]{B}(p,2\alpha)\isotope[4]{He}$ yield is greater in direct-irradiation type of experiments.  For these we need a reaction occuring in parallel with a product that escapes the plasma unperturbed, such as neutrons and photons.  

To strengthen identification of the originating reaction, a neutron or photon produced with definite energy is preferable.  Exothermic reactions ($Q>0$) are therefore better candidates, since we can expect the neutron or photon spectrum to peak at nonzero kinetic energy.  Of the primary reactions \tab{primaryreactions}, only two satisfy these conditions, $\isotope[11]{B}(p,\gamma)\isotope[12]{C}$ and $\isotope[10]{B}(p,\gamma)\isotope[11]{C}$, both emitting photons.  The cross section data for both reactions is very limited.  For the $\isotope[10]{B}(p,\gamma)\isotope[11]{C}$ reaction especially, the trend in the available data suggests our calculations here may significantly underestimate the photon yield.  Of the secondary reactions, three satisfy these conditions, $\isotope[11]{B}(\alpha,n)\isotope[14]{N}$, $\isotope[10]{B}(\alpha,n)\isotope[13]{N}$ and $\isotope[14]{N}(\alpha,\gamma)\isotope[18]{F}$.  

As several of the candidate reactions have limited cross section data available and even more limited data on the spectrum of the outgoing neutron or photon, it would be a reasonable first step to verify yields with beam-target type of experiments.  Since the target remains intact, we can measure a wider variety of reaction products, in particular unstable nuclides with half-lives much greater than the experiment duration.  Several of the reactions selected by the previous approach also yield unstable nuclei, specifically $\isotope[11]{C}$, $\isotope[13]{N}$ and $\isotope[18]{F}$ with half lives on the order of $10^3$ s.  Other unstable nuclides produced are listed in \tab{unstablenuclides}.  The significantly differing half lives make identification by reactivity straightforward with a Geiger counter placed near the target, though $\isotope[14]{C}$ probably has too long a half life and $\isotope[10]{C}$ too short a half life for reliable identification.  $\isotope[7]{Be}$ is undetectable by this means, but arises from a reaction without particular interest in this study. 

\begin{table}
\begin{tabular}{l|c|c}
Nuclide & Half life  [s] & Decay mode \\\hline
$\isotope[7]{Be}$ & $4.60\times 10^6$ & $\epsilon$ \\
$\isotope[10]{C}$ & $19.3$ & $\beta^+$ \\
$\isotope[11]{C}$ & $1.22\times 10^3$ & $\beta^+$ \\
$\isotope[13]{N}$ & $598$ & $\beta^+$ \\
$\isotope[14]{C}$ & $1.81\times 10^{11}$ & $\beta^-$ \\
$\isotope[14]{O}$ & $70.6$ & $\beta^+$ \\
$\isotope[17]{F}$ & $64.5$ & $\beta^+$ \\
$\isotope[18]{F}$ & $6.59\times 10^3$ & $\beta^+$ \\ \hline
\end{tabular}
\caption{\label{tab:unstablenuclides} Half lives and decay modes of unstable nuclei produced by reactions in \tabs{primaryreactions}{secondaryreactions} }
\end{table}

Now synthesizing and narrowing the list of promising reactions, two of the photon- and neutron-producing reactions stand out in utility.  First, $\isotope[11]{B}(p,\gamma)\isotope[12]{C}$ is promising to directly correlate with the reaction $\isotope[11]{B}(p,2\alpha)\isotope[4]{He}$, because it has the same initial state and produces a photon with energy significantly above most other products.  The $\sim 100\times$ larger cross section for $\isotope[10]{B}(p,\gamma)\isotope[11]{C}$ make it a practical proxy in the near term, though it has a different initial state, which introduces additional uncertainty.  Ultimately the yields of these two reactions will determine which is more useful in the experiment.

The second $\isotope[10]{B}(\alpha,n)\isotope[13]{N}$ could help verify the $\alpha$ yield.  Since this reaction is the isospin partner of $\isotope[10]{B}(\alpha,p)\isotope[13]{C}$, the cross sections are very similar in magnitude and energy dependence and its measurement would verify the role of proton-recycling from secondary reactions, proposed as an important mechanism in enhancing $\isotope[11]{B}(p,2\alpha)\isotope[4]{He}$ yields in some experiments \cite{labaune2016laser}.  Moreover, $\isotope[13]{N}$ is unstable and its yield can be measured independently in beam-target experiments.  A small drawback to this reaction is that the cross section decreases sharply below CM energy of 3 MeV.  The majority of $\alpha$s produced by  $\isotope[11]{B}(p,2\alpha)\isotope[4]{He}$ should have energy above this threshold, but $\alpha$s also lose energy quickly in a cold medium and the yield is expected to be small. 

Of the remaining two, (a)  there is unfortunately no cross section data for $\isotope[14]{N}(\alpha,\gamma)\isotope[18]{F}$, though it should produce a peak in the photon spectrum, and (b) the neutron produced in $\isotope[11]{B}(\alpha,n)\isotope[14]{N}$ is likely have very little kinetic energy, making detection difficult.

Of the radioactive nuclides, $\isotope[11]{C}$ has already proven its utility in beam-target experiments providing a significantly higher-confidence estimate of the $\alpha$ yield than a direct measurement of the $\alpha$s by CR-39 \cite{McCary2022}.  $\isotope[18]{F}$ has been detected \cite{labaune2016laser,McCary2022} but the absence of a cross section for its production limits the information gained.  More surprisingly, $\isotope[13]{N}$ has not been detected, probably due to a combination of shorter halflife and low yield making it difficult to distinguish from the $\isotope[11]{C}$ signal.  Since better data exists for the $\isotope[13]{N}$ production cross section and it is generated almost entirely by $\alpha$ scattering, we consider it the most important candidate for future experiments as a means to help verify the $\alpha$ yield.  $\isotope[14]{O}$ and $\isotope[17]{F}$ are of little interest: in both production processes, $Q<0$ so the neutron released is not a potential diagnostic in direct irradiation experiments.

\section{Yield equations}

Having winnowed the set of interesting reactions based on general criteria for good diagnostics of the fusion dynamics, we now evaluate the yields of the various products.  

The challenge in deriving analytic expressions for the yield expression is two-fold.  First, the ion momentum distribution function generally varies from shot to shot due to variations in the laser.  Therefore the ion distribution should be measured on each shot as well as possible and used in predictions.  We address modeling related to incomplete measurement in \ssec{modelresults}.
  
Second, the ion momentum distribution is heterogeneous, with two or more distinguishable populations.  Short-pulse high-intensity lasers deposit energy into a region 10-100 $\mu$m in radius from the focal spot, which we call the directly-irradiated (DI) volume.  In this volume, electrons gain many MeV of energy and ions are likely to have more isotropic momentum distributions.  Fast electrons pushed through and out of this volume, largely in the laser beam direction, can create large magnitude electrostatic fields that accelerate ions out of the DI volume.  The precise ion acceleration mechanism and shape of the resulting ion spectrum depend on the thickness of the target.  To maximize fusion yields, we assume the target is thick, that is greater than the stopping range of these high-energy ions, so that the probability of undergoing fusion is saturated.  The fast ions are more likely to collide with at-rest ions much deeper in the bulk of the target, and the reaction kinematics are essentially those of beam fusion.  We model the yield from each of these regions and ion populations separately, addressing reactions in the DI volume first and the beam fusion reactions second.

The dichotomy between the DI volume and the beam fusion region is artificial and the plasma will certainly contain some transitional regions.  Given the dynamics described so far, these transitional regions are likely to contain electrons of intermediate kinetic energy $1\,\mathrm{eV}\ll \langle E_e\rangle\ll 1\,\mathrm{MeV}$, both low energy ions and some beam ions and particle number densities similar to the initial state.  In this region, ion stopping is reduced compared to the cold limit and the fusion probability is similar to the beam fusion limit.  Thus, the contribution to the yield can be thought of as a correction to the beam fusion yield since the length of such transitional regions is much less than the stopping range of the fast ions passing through.

The starting point to derive the yield is a classical expression for the total number of particles of type $A$ produced in a 2-body collision,
\begin{align}\label{eq:Ystarting}
Y_A=\int f_1(\vec x,\vec p_1,t)f_2(\vec x,\vec p_2,t)|\vec v_1-\vec v_2|\sigma_Ad^3x dt \frac{d^3p_1d^3p_2}{(2\pi)^3(2\pi)^3}.
\end{align}
Here $f_i(\vec x,\vec p_i,t)$ is the distribution function describing the probability of finding particle $i=1,2$ with momentum $\vec p+d p$ in the volume element $\vec x+dx$ at time $t$, $\sigma$ is the cross section, and $|\vec v_1-\vec v_2|$ is the relative speed of the incident particles.  We integrate over all initial particle momenta, all final states, and all space and time for the reaction to occur.  For the reactions of interest here, sufficiently complete cross sections differential in solid angle are generally not available, and cross sections for the process exhibit resonances of nuclei with different quantum numbers, suggesting that the angular dependence will have a strong energy dependence, which we will not attempt to model here.  We can focus on the total yield as the most relevant observable for both applications and practically available measurements from recent experiments.

We consider two models for our yield calculations, corresponding to the direct-irradiation and beam-target experiments.  

\subsection{Direct irradiation}
The laser deposits a large amount of energy in the fusion target and reactions occur within the 10s of picosecond timescale that the ions are heated but before the target expands and its density drops.  Although neither the electron nor the ion population can equilibrate in this short time, experimental ion spectra are frequently fit by a Maxwellian distribution, $dN/dE\propto e^{-\beta E}$.  The parameter $\beta$ is an inverse energy scale that characterizes the mean kinetic energy per particle of the distribution $\langle E\rangle/N=\beta^{-1}$.  We stress that local thermal equilibrium is certainly not achieved and we do not assume equilibrium distributions.  The Maxwellian fit to experimental spectra is a phenomenological choice, enabling simple quantitative comparison between shots and facilities. 

Maxwellian ion distributions is a strong simplifying approximation: kinetic simulations of short-pulse laser-target interaction have found the ion distributions can have significantly higher numbers of high-energy ions $E_i>\beta^{-1}$ than expected from a Maxwellian distribution.  To account for the excess of high-energy $E>\beta^{-1}$ ions, the single Maxwellian model can be improved by introducing a second Maxwellian distribution of smaller $\beta$.  This second population often corresponds to the beam population that has significant directionality and is less likely to react with the larger $\beta$ population, being accelerated out of the DI region by plasma fields on its boundary.  Even so, the contribution is computed easily since the yield is linear in the distribution function and the yield in the double Maxwellian case can be derived by summing four yields corresponding to the four combinations of the two ion species' two $\beta$ values.  Therefore for simplicity and clarity here, we use the single Maxwellian.

Since the electrons have MeV-scale kinetic energy in the DI region, their stopping power is significantly reduced.  While the high-intensity laser can drive large, short-lived, local increases in the electron density, the ion density varies from the initial value by a factor much less than 1, at least until the target expands significnatly on the 10s of picosecond timescale.  Consequently, the proton and $\alpha$ stopping ranges ($>100\,\mu$m) are certainly larger than the radius of the DI region ($\lesssim 100\,\mu$m), and we consider that the ion energy losses are negligible for the duration in the DI region.

Given these conditions, the yield of a given process is straightforwardly derived from \eq{Ystarting}.  The reaction volume is a few times larger than the focal volume but is generally not known precisely.  Reactions will continue as long as the plasma remains relatively dense, up to several picoseconds, though this plasma ``confinement time'' as it is sometimes known is not well-determined either.   We therefore consider the yield per unit volume per unit time for a $2\to A+X$ reaction for nonrelativistic ions with Maxwellian distributions,
\begin{align}\label{eq:Ythermalrate}
\frac{dY_A}{d^3xdt}&=\frac{n_1n_2}{\pi}\left(\frac{2\beta_1m_1\beta_2m_2}{m_1\beta_1+m_2\beta_2}\right)^{\!3/2}\frac{1}{\beta_rm_r}\int_0^\infty \! \mathcal{Y}(m_r,\beta_r,\mu,\nu;v)\,\sigma_A\!\Big(\frac{m_r}{2}v^2\Big)\,v\,dv\\ \notag 
\mathcal{Y}(m_r,\beta_r,\mu,\nu;v)&=e^{-\mu v^2}\left(e^{y^2_v}\big(2y^2_v+1\big)\frac{\sqrt{\pi}}{2}\mathrm{Erf}\!(y_v)+y_v\right), \qquad
y_v= \frac{\beta_rm_rv}{\sqrt{2\nu}} 
\end{align}
where the parameters $n_i,m_i,\beta_i$ for $i=1,2$ are the number densities, masses and inverse mean kinetic energy of the two ion species.  The remaining parameters are the reduced mass, $m_r$, difference of inverse mean kinetic energy and combinations thereof,
\begin{align}\label{eq:Yparamsdefns}
m_{r}=\frac{m_1m_2}{m_1+m_2}, \qquad 
\beta_r=\beta_1-\beta_2, \qquad
\nu=\beta_1m_1+\beta_2m_2, \qquad
\mu=\frac{m_r^2}{2}\left(\frac{\beta_1}{m_1}+\frac{\beta_2}{m_2}\right) .
\end{align}
The result is even in $\beta_r$ as it must be since the choice of labels is arbitrary and the yield should always be positive.  The integration variable corresponds to the magnitude of the relative velocity of the two ions.  The integration will be performed numerically to use experimental data for the cross section $\sigma_A(E_{cm})$, which is a function of the CM energy.  Erf$(z)$ is the usual error function with the normalization defined by
\begin{align}\label{eq:erfdefn}
\mathrm{Erf}(z)=\frac{2}{\sqrt{\pi}}\int_0^ze^{-u^2}du.
\end{align}
The limit of equal mean kinetic energies is simplifies the result considerably to
\begin{align}\label{eq:YthermalequalT}
\frac{dY_A}{d^3xdt}=\frac{2n_1n_2}{\pi\sqrt{\beta(m_1+m_2)}}\int_0^\infty \!dy e^{-y^2/2} y^2\sigma_A\!\Big(\frac{y}{2\beta}\Big)
\end{align}

Considering our interest in particular reactions as \emph{in-situ} diagnostics of the  $\isotope[11]{B}(p,2\alpha)\isotope[4]{He}$ reaction, we introduce ratios of yields to eliminate experimental unknowns.  For reactions with the same initial state, e.g. p-$\isotope[11]{B}$ scattering, all the prefactors in \eq{Ythermalrate} cancel.  For example, to use the $\isotope[11]{B}(p,\gamma)\isotope[12]{C}$ reaction as a diagnostic on $\isotope[11]{B}(p,2\alpha)\isotope[4]{He}$, we might consider the ratio 
\begin{align}\label{eq:Y12CgammabyYalpha}
\frac{dY_\gamma}{dY_\alpha}=\frac{\int_0^\infty \!dv \mathcal{Y}(m_r,\beta_r,\mu,\nu;v)\,v\,\sigma_{pB\rightarrow \isotope[12]{C}\gamma}\!\Big(\frac{m_r}{2}v^2\Big)}{3\int_0^\infty \!dv \mathcal{Y}(m_r,\beta_r,\mu,\nu;v)\,v\,\sigma_{pB\rightarrow 3\alpha}\!\Big(\frac{m_r}{2}v^2\Big)},
\end{align}
in which all the mass- and $\beta$-dependent parameters are identical in the numerator and denominator.  Only the cross sections differ.  For $2\to 2$ reactions such as considered here, the spectra of produced neutrons and photons have been computed semianalytically showing that their widths and small shifts in the peak depend on the momentum distribution of scattering ions \cite{appelbe2011production}.  This allows the (approximate) $\beta$ parameters of the ions to be retrieved by fitting spectra of the measured neutron or photon.  Since the reactions have exactly the same initial state, potentially large scaling factors such as volume and time must be the same.  Thus this yield ratio depends only on the mean kinetic energies of the two ion species.  Considered as a function of these two energy scales, the ratio manifests the difference in the energy dependence of the cross sections, though less so than the beam-target experiments described below.  In \eq{Y12CgammabyYalpha}, the factor 3 has been included in the numerator to count the total number of $\alpha$s produced for each $\isotope[11]{B}(p,2\alpha)\isotope[4]{He}$ reaction.  With this yield ratio, the number of $\isotope[11]{B}(p,2\alpha)\isotope[4]{He}$ reactions is recovered by multiplying by the measured yield of photons identified as arising from this reaction.

Another reaction of interest for diagnostics is $\isotope[10]{B}(p,\gamma)\isotope[11]{C}$, which differs from $\isotope[11]{B}(p,2\alpha)\isotope[4]{He}$ in the isotope of boron in the initial state.  As a consequence, some prefactors remain in the yield ratio,
\begin{align}\label{eq:Y11CgammabyYalpha}
\frac{dY_\gamma}{dY_\alpha}=
\frac{n_{10}}{n_{11}}\left(\frac{1+\frac{m_p}{m_{11}}\frac{T_{11}}{T_p}}{1+\frac{m_p}{m_{10}}\frac{T_{10}}{T_p}}\right)^{\!3/2}\frac{T_{r10}}{T_{r11}}\frac{m_{r11}}{m_{r10}}
\frac{\int_0^\infty \!dv \mathcal{Y}(m_{r11},T_{r11},\mu_{11},\nu_{11};v)\,v\,\sigma_{pB\rightarrow \isotope[11]{C}\gamma}\!\Big(\frac{m_r}{2}v^2\Big)}{3\int_0^\infty \!dv \mathcal{Y}(m_{r10},T_{r10},\mu_{10},\nu_{10};v)\,v\,\sigma_{pB\rightarrow 3\alpha}\!\Big(\frac{m_r}{2}v^2\Big)}.
\end{align}
The subscripts for boron parameters have been shortened to the isotope number for clarity.
The measured constants in the prefactor, such as masses, are no trouble, but for this ratio to be useful we must argue that the ratio of densities remains nearly constant during the relevant period of plasma evolution. Since the charge is the same and the masses differ only by 10\%, we suppose significant separation of isotopes from the initially uniform mixture can only develop slowly, on the same time scale (or longer) that the plasma expands and diffuses into free space.  Note that dynamically, we expect the mean kinetic energy of the boron ions and protons to be similar, making the prefactor in parentheses close to 1. The remaining ratio of reduced kinetic energies is expected to be near unity for the same reason.

Another useful yield ratio could be $\isotope[10]{B}(\alpha,p)\isotope[13]{C}$ relative to
$\isotope[10]{B}(\alpha,n)\isotope[13]{N}$.  The ratio would cancel dynamical unknowns such as the density of $\alpha$s.  The other proton-producing secondary reaction $\isotope[11]{B}(\alpha,p)\isotope[14]{C}$ could be added to the ratio to completely determine the secondary proton production, though the same remarks as above would apply to the prefactor.   A measurement of the neutrons produced from $\isotope[10]{B}(\alpha,n)\isotope[13]{N}$ constrains the number of protons able to be recycled into the $\isotope[11]{B}(p,2\alpha)\isotope[4]{He}$ reaction.  In this case though the input is the $\alpha$ spectrum derived from all primary reactions, is a complicated function of energy and is expected to vary significantly as a function of mean ion kinetic energy.  We consider its derivation beyond the scope of this paper.  

\subsection{Beam-target}
The beam-target experiment involves simpler kinematics.  In the frame with the target material at rest, the center of mass energy is
\begin{align}
E_{cm}=\frac{m_r}{2}\vec v_b^2=\frac{m_r}{m_b}E_b,
\end{align}
with the reduced mass given above by \eq{Yparamsdefns}.  The $b$ subscript indicates a particle from the beam and the $t$ subscript indicates a particle in the target.  For $pB$ scattering $m_{r}/m_b\simeq 1.1$.
The momentum integral then only runs over the proton distribution. The target particle distribution function is nonzero only in the spatial region of the target material, and integrating over the beam axis and time convolves the projectile beam with the target distribution.

In standard beam-target experiments in order to maximize exposure, the target is placed on or adjacent to the anticipated axis of the ion beam.  In the TPW experiment of particular interest, we were able to verify that the target material contained the cone of highest ion flux, which subtended an opening angle $\theta\lesssim 0.3$.  We therefore assume that the transverse momentum of the beam is small relative to longitudinal momentum.  These together imply we can reduce the beam momentum integral to the longitudinal momentum only and integrate the transverse position dependence into a 1-dimensional beam distribution function,
\begin{align}\label{eq:Ybeam1d}
Y_A&\simeq\int_{V_t}d^3x \int_{-\infty}^\infty dt \int\frac{dp_z}{2\pi} f_b(\vec x,p_z,t)n_t(\vec x)|\vec v_b|\sigma_{\!A}(E_{cm})  \\ \notag
&=n_t\int_0^L dz \int_{-\infty}^\infty \!dt \int_0^\infty\frac{dp_z}{2\pi} v_z\,\sigma_{\!A}(E_{cm})\, f_b(z,p_z,t)
\end{align}
where $V_t$ signifies the volume of the target.  In this expression the longitudinal coordinate can also be considered as parameterizing the distance along the on-average straight-line trajectory; trajectories diverging from the beam axis would make a small geometric correction due to exiting through the side of the target rather than the opposite end. The constant density of the target has been taken outside the integral and the target length defined as $L$.  

Due to energy loss in the target, the beam distribution evolves as its propagates through the target.  First, as a limiting model, we compute the yield neglecting the beam energy loss.  This case also clarifies the dynamics in the subsequent derivation that includes stopping.  The beam distribution function remains constant in the absence of stopping, so the convolution yields the length of the target times the spatial length scale of the beam divided by the longitudinal velocity, i.e. the length of the beam multiplying the traverse time and a numerical factor depending on the longitudinal profile of the beam.  Then the yield can be written simply
\begin{align}\label{eq:nolossyield}
Y_A^{\emptyset}&\simeq n_tL \int\frac{dE_b}{2\pi}\, \frac{dN_b}{dE_b}\,\sigma_A\!\left(\frac{m_r}{m_b}E_b\right).
\end{align}
The beam distribution function $f_b$ has been reduced to its energy dependence $dN_b/dE$.

The importance of stopping is seen by comparing the stopping range to the target dimensions.  The stopping range is defined
\begin{align}\label{eq:stoppingrange}
z_s=\int_0^{E_0}\left(\frac{dE}{dx}(E')\right)^{\!-1}dE', 
\end{align}
where $E_0$ is the initial energy of the ion before interacting with the target and $dE/dx$ from data is conventionally positive and the expected minus sign is compensated by the flipping the limits on the integral.  Note that $dE/dx$ is frequently given in units of energy/(mass density) or energy/(number density) so that one multiplies by the density of the medium to obtain the energy loss in units of energy/length.  

The target temperature is more difficult to estimate in direct irradiation experiments.  As the laser energy is absorbed within the first few 10s of microns of the target (at most), the bulk is only heated by ions and electrons accelerated out of the laser-heated region.  Ions dominate the energy transfer to the bulk; electrons have very low (a few MeV cm$^2$/g) stopping power in the few-MeV energy range compared to ions.  With similar estimates for the total energy of laser-accelerated ions as in the preceding paragraph, the average energy transferred is 10-100 eV per electron, orders of magnitude higher because the volume into which it is deposited is orders of magnitude smaller $\sim (0.1\mathrm{mm})^3$.  The temperature-dependent correction to ion stopping would be non-negligible in this case.  For this reason, in yield calculations below, we compare zero-temperature stopping to finite-temperature stopping.

For a zero-temperature boron target and ion energies representative of the higher end of the expected distributions, SRIM predicts the stopping range of a 20 MeV proton as 2 mm and 8 MeV $\alpha$ as 38 micron.  However, target can be heated by both the ion beam and the even higher energy electrons that are accelerated out of the ion source by the laser driver.  Using the fact that the stopping range is less than the target length even for the highest energy ions, the total energy deposited is just the total energy of the beam that enters the target.  Even for the relatively high energy ions obtained from the Texas Petawatt, the total ion beam energy transferred to the target is at most $\sim 10\%$ of the laser energy.  For the upper limit on the Texas Petawatt, 10 J deposited into a hemisphere of radius equal to the 2mm stopping range, the specific heats of boron and boron nitride imply a temperature change $\Delta T\simeq 280$ K$\simeq 2.4$-$2.6\times 10^{-2}$ eV).  Without direct measurements of the electron spectrum emitted by the ion source, we resort to an estimate.  While experiments and simulations of ion acceleration suggest that electrons absorb a similar amount of energy from the laser-plasma interaction as the ions, the electrons are less efficient at depositing energy in the target.  Therefore an estimated upper bound on the deposition of energy in the target by electrons is 10 J.  Carbon and other heavy ions that may come from the ion sources carry equal or less energy than the protons and in any case arrive later.  Thus our best estimate for the temperature of the target remains $\Delta T\lesssim 5\times 10^{-2}$ eV.  This estimate, much less than the work function ($\sim eV$) of the target material, is consistent with the target's survival of the interaction.

\begin{figure}
\includegraphics[width=0.48\textwidth]{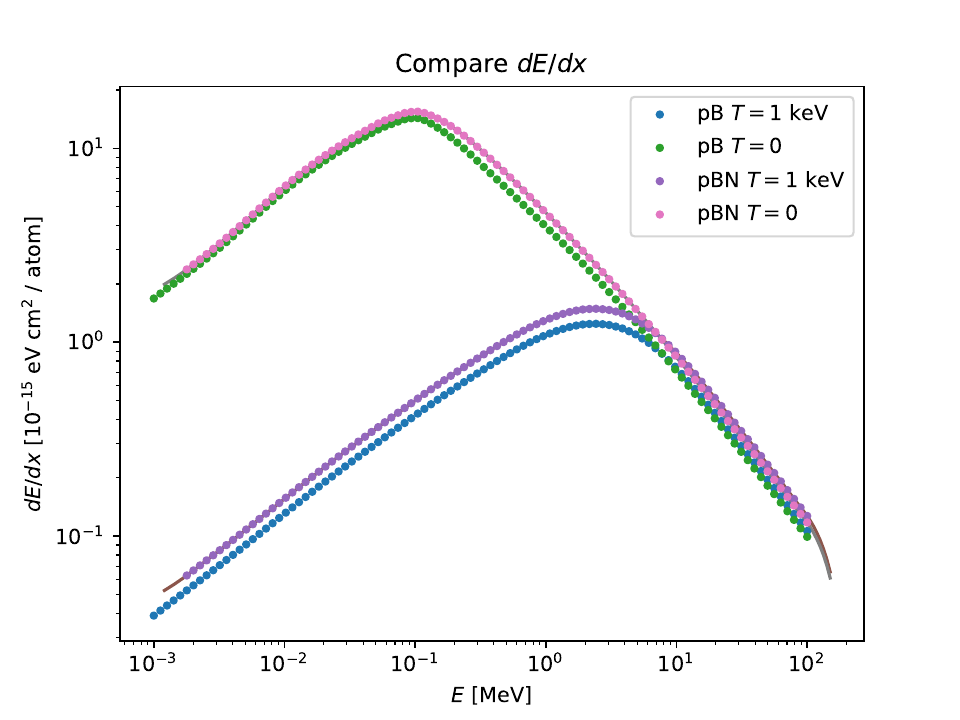}
\includegraphics[width=0.48\textwidth]{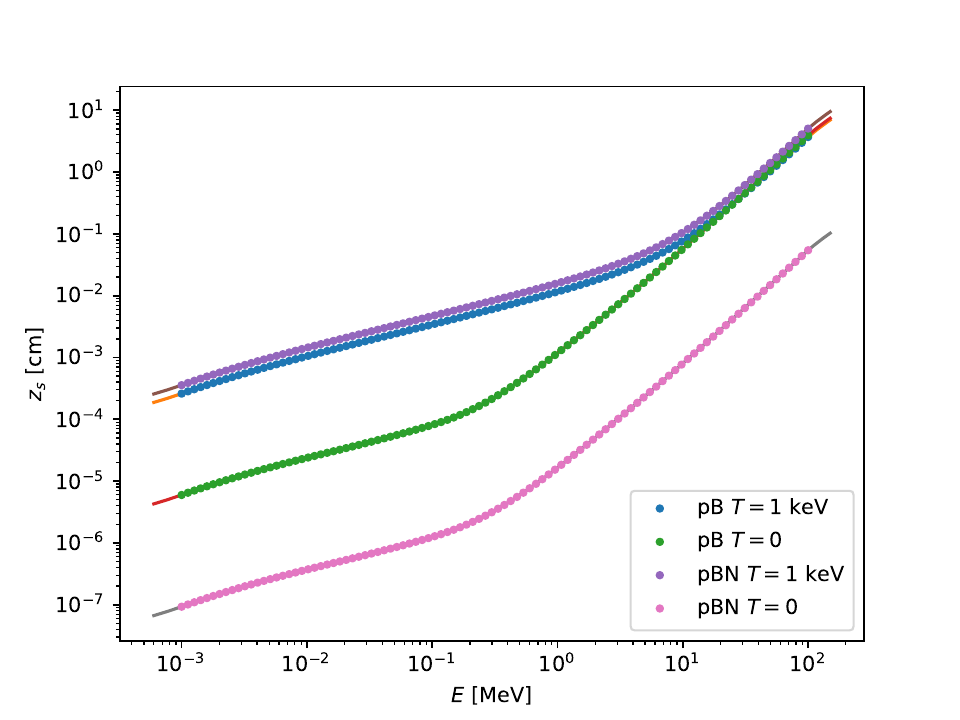}
\caption{\label{fig:dEdx} Left: Stopping power, $dE/dx$, for protons in pure boron and boron-nitride, cold and warm $T=1$ keV thick targets.  Right: Resulting range $z_s$ \eq{stoppingrange}. }
\end{figure}

Nevertheless, for reference and comparison, \fig{dEdx} shows the stopping power and stopping range \eq{stoppingrange} for both cold and high-temperature ($T=1$ keV) boron and boron-nitride.  This unphysically high target temperature is chosen to exhibit its negligible impact on the yields for the processes of interest.  The stopping power data are obtained from calculations using the enhanced RPA-LDA (eRPA-LDA) model of Gu, Mehlhorn, and Golovkin \cite{mehlhorn1981finite,gu2022}.  Stopping ranges for lower energy ions are shorter, and ranges are generally less than the typical length ($\lesssim$ cm) of the targets.  The highest energy protons ($E_p\gtrsim 20$ MeV) have a stopping ranges equal or greater than the target length, but their number and hence contribution is smaller by an order of magnitude or more.  Neglecting this not-quite-stopped component therefore amounts to an error of $\sim 10\%$ or less, smaller than the error propagated from the cross section and certainly smaller than the error due to the limited energy range of the cross section data.  Therefore, to our working accuracy, the target can be considered ``thick'' in that almost all particles in the beam will be stopped.  

The conventional definition of the ``thick-target yield'' for a monoenergetic input is
\begin{align}
I_A(E_0)=\int^{E_0}_0\left(\frac{dE}{dx}(E')\right)^{\!-1}\sigma_A\!\big(\frac{m_r}{m_b}E'\big)dE'
\end{align}
where $dE/dx$ from data is conventionally positive and the expected minus sign is compensated by the flipping the limits on the integral.  The density factor in converting tabulated $dE/dx$ data into energy loss per unit length cancels with the density of target nuclei in the yield.  Note that the integration can effectively be restricted to the energy range where the cross section is nonnegligible.  Since most of the cross sections have thresholds of order 1 MeV, \fig{dEdx} shows that finite temperature corrections to stopping matter only for $T\gtrsim 1$ keV.  Raising the target temperature increases the projectile energy at which $dE/dx$ achieves its maximum, but 1 keV is already much higher than can be realistically achieved in the beam-target experiment. 

\begin{figure}
\includegraphics[width=0.48\textwidth]{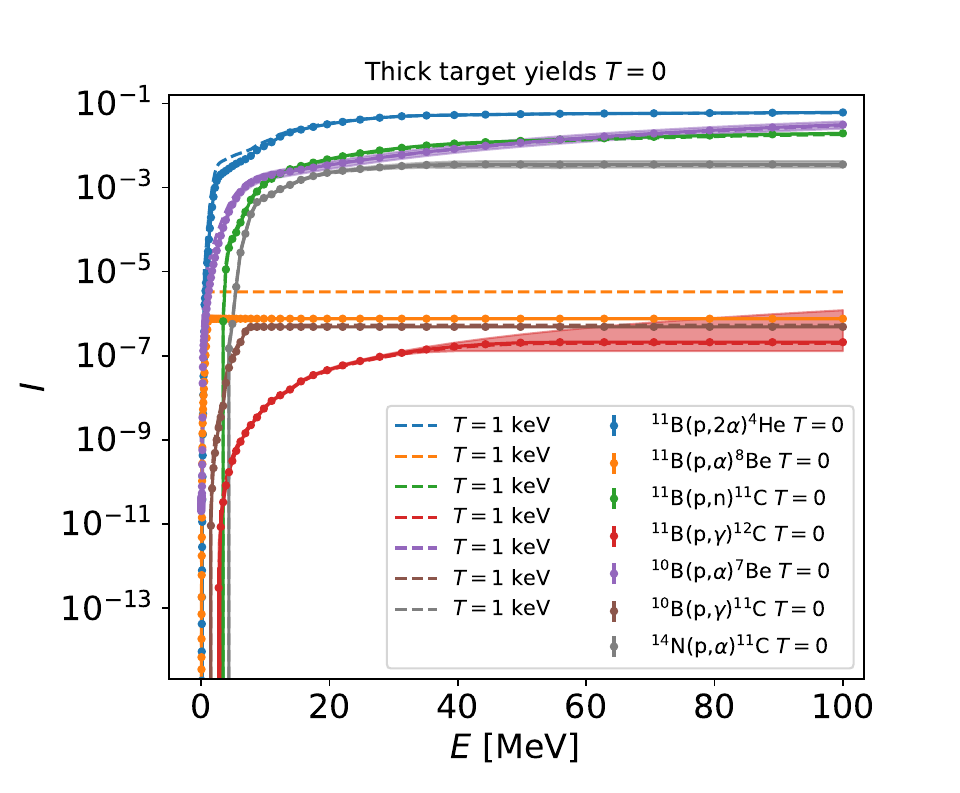}
\caption{\label{fig:Ithick} Thick target yields for the reactions in \tab{primaryreactions}. Solid bands show the error propagated from the cross section.  Error bars, where visible, present the numerical error. }
\end{figure}

Thick target yields for all of the primary reactions in \fig{Ithick} show these properties.  $\isotope[11]{B}(p,2\alpha)\isotope[4]{He}$ and $\isotope[11]{B}(p,\alpha)\isotope[8]{Be}$ display the greatest sensitivity to the target temperature because the $\isotope[11]{B}(p,2\alpha)\isotope[4]{He}$ cross section is largest and the cross section for $\isotope[11]{B}(p,\alpha)\isotope[8]{Be}$ is only available for CM energy $\lesssim 1 MeV$, where the $dE/dx$ curves differ the most.

The total yield of the product nucleus $A$ is obtained by integrating the thick target yield over the beam, weighted by the beam energy distribution $dN/dE_b$,
\begin{align}
Y_A^{\rm tt}=\int_0^{\infty}\frac{dN}{dE_b}I(E_b) dE_b.
\end{align}
Yield ratios in the beam-fusion geometry, as in direct irradiation experiments, analytically eliminate dependence on geometric factors in the yield, such as target length and density.  Less obviously, the overall normalization of the beam energy distribution also cancels in the ratio, since one could easily write $dN/dE=N_bf(E)$ where $N_b$ is the total number of particles (that interact with the target) and $f(E)$ is a normalized probability distribution for the ion energy.   For example, the ratio of $\gamma$s from  $\isotope[11]{B}(p,\gamma)\isotope[12]{C}$ to $\alpha$s from $\isotope[11]{B}(p,2\alpha)\isotope[4]{He}$, is
\begin{align}  
\frac{Y_\gamma}{Y_\alpha}=\frac{\int_0^{\infty}f(E_b)\int^{E_b}_0\left(\frac{dE}{dx}(E')\right)^{\!-1}\sigma_{pB\to^{12}\mathrm{C}\gamma}\!\big(\frac{m_r}{m_b}E'\big)dE'dE_b}{\int_0^{\infty}f(E_b)\int^{E_b}_0\left(\frac{dE}{dx}(E')\right)^{\!-1}\sigma_{pB\to 3\alpha}\!\big(\frac{m_r}{m_b}E'\big)dE'dE_b}
\end{align}
with the same stopping power $dE/dx$ and normalized proton spectrum.
 Removing this dependence on the total number in the beam significantly reduces uncertainty in practice given the available on-shot beam measurements.  

\begin{figure}
\includegraphics[width=0.5\textwidth]{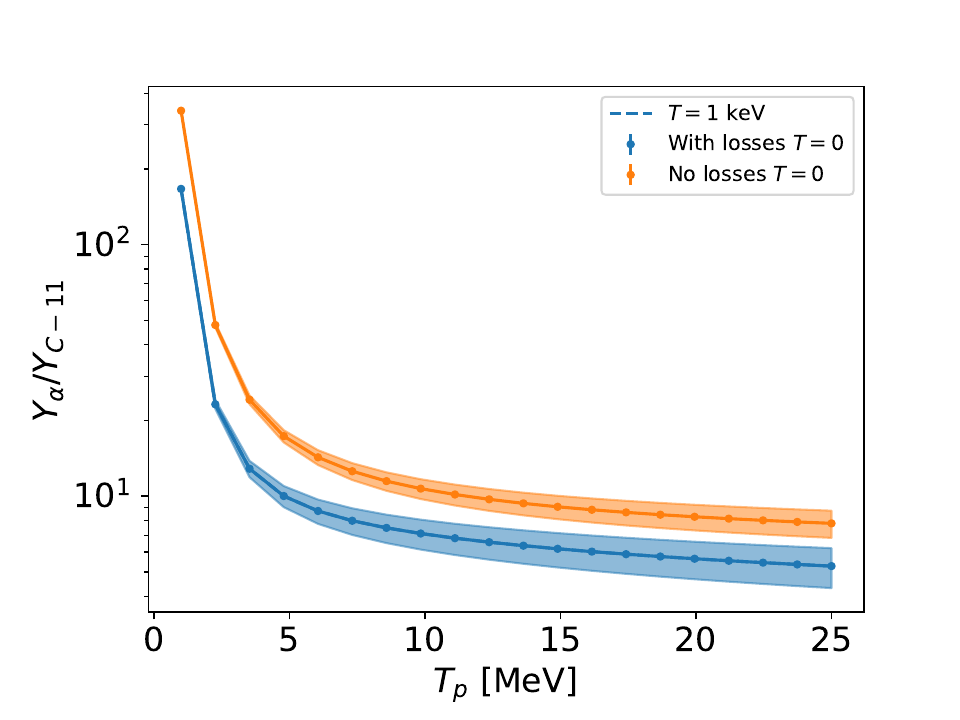}
\caption{\label{fig:alphaby11Chotvcoldlosses} Yield ratio of $\alpha$ to $\isotope[11]{C}$ for a normalized Maxwellian input proton spectrum in boron, comparing the yield with and without beam energy losses in the target.  Target temperature makes a negligible difference to the energy loss for proton energies in this regime, and the $T=1$ keV curve lies on top of the $T=0$. Solid bands show the error propagated from the cross section.  Error bars, where visible, present the numerical error.}
\end{figure}

The yield ratio retains important information of the beam energy distribution.  As seen in \fig{sigmacomp}, different reactions have different thresholds, collision energies where the cross section approaches its maximum usually in the 0.1-1 b range.  The yield ratio is greatly enhanced in case the beam energy distribution reaches the threshold of one reaction but not the other.  Since laser-driven ion beams generally have a broad and decreasing energy distribution at low energy, the typical case is that a beam may contain ions of sufficient energy for a reaction with a low threshold but not for a reaction with a higher threshold.
Thus for example the $\isotope[11]{B}(p,2\alpha)\isotope[4]{He}$ has a peak cross section around 650 keV whereas the $\isotope[11]{B}(p,n)\isotope[11]{C}$ has a peak around 7 MeV, so that a Maxwellian distribution with $\beta^{-1}\simeq 0.5$-5 MeV would yield significant $\alpha$ particles but not $\isotope[11]{C}$.  This effect is demonstrated in \fig{alphaby11Chotvcoldlosses}.

\section{Modeling and results}\label{sec:modelresults}

Some modeling and assumptions have already been established in setting up expressions for the yield.  We now discuss the details and quantitative inputs to the models.

As seen above in \tabs{primaryreactions}{secondaryreactions} and \fig{sigmacomp}, data for the cross sections of interest is available only for limited ranges of CM energy.  The available data and comparison with analogous reactions suggest that the cross sections may have similar values across a wider range of CM energies.  However to avoid undue speculation and modeling, we assume the cross section vanishes quickly outside the available data range.  Considering also the ion mean kinetic energy not likely exceeding $\sim 10$ MeV, the numerical results for the yields are likely under estimates by a factor of a few, but not more than ten.

\subsection{Direct irradiation}

The yield ratio eliminates dependence on local, dynamic quantities, including effective reaction volume, confinement time and the densities.  We need both the absolute kinetic energy scale and the relative kinetic energies of the two ion species.  The absolute energy scale is determined by how efficientlly laser energy is transferred the plasma, which in turn generally depends on laser properties, such as total pulse energy, pulse length (if it is greater than ps-scale), and contrast.  For comparison between facilities we scan the absolute kinetic energy scale, using the proton mean kinetic energy as the reference.  For intensities up to $10^{22}$ W/cm$^2$, we expect ion kinetic energies inside the target to be MeV-scale as the typical momentum obtained from a cycle of the laser field and so also from plasma-generated electrostatic fields.

\begin{figure}
\includegraphics[width=0.48\textwidth]{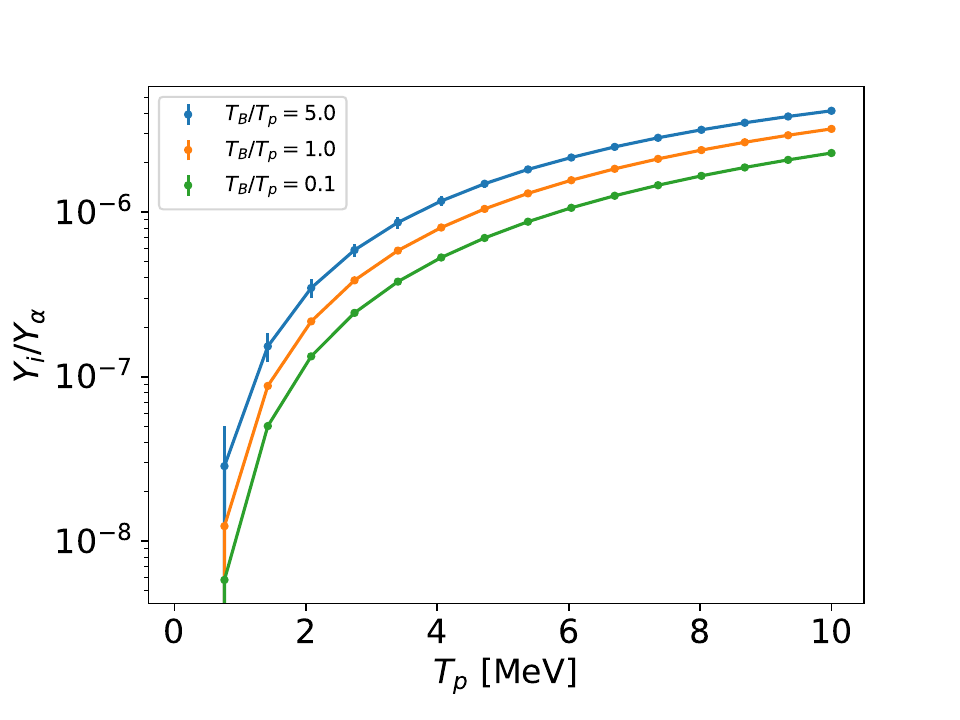}
\includegraphics[width=0.48\textwidth]{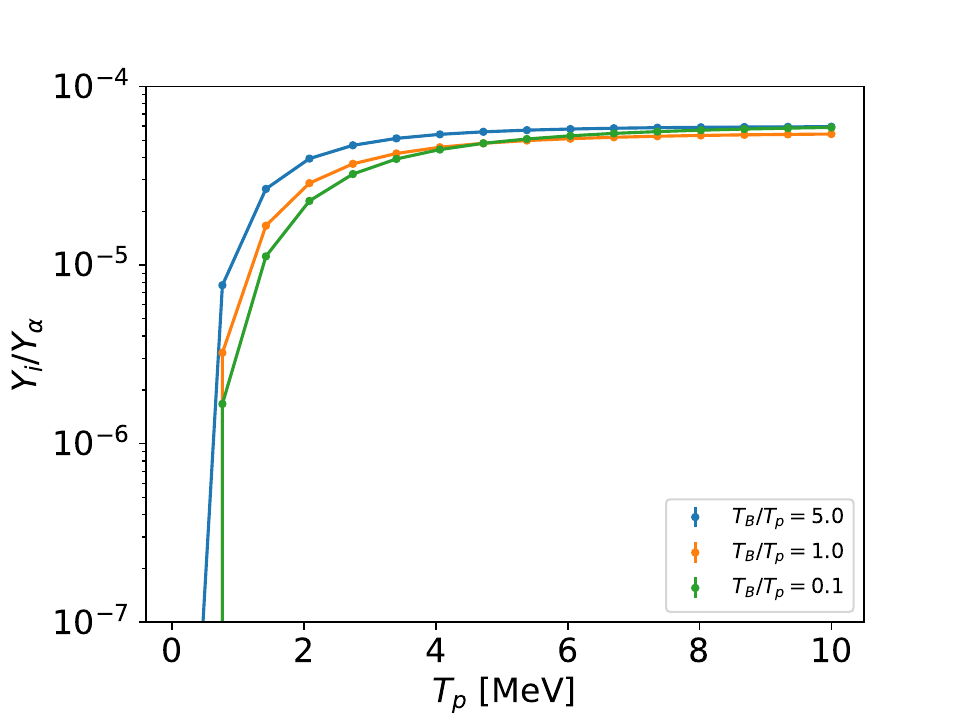}
\caption{\label{fig:DIyieldratios} Ratio of photon ($\gamma$) yield to $\alpha$ yield in a quasi-thermal plasma as a function of proton mean kinetic energy.  At left for the $\isotope[11]{B}(p,\gamma)\isotope[12]{C}$ reaction and at right for the $\isotope[10]{B}(p,\gamma)\isotope[11]{C}$ reaction.  The yield increases rapidly as the mean kinetic energy nears the CM energy corresponding to the threshold for the cross section (cf \fig{sigmacomp}) and then plateaus.  Different curves correspond to different boron to proton kinetic energy, showing that the yield is only sensitive to the relative kinetic energy if the boron mean kinetic energy is much less than the proton mean kinetic energy.  Error bars, where visible, present the numerical error.}
\end{figure}

The relative kinetic energy can be estimated from kinematics.  For the same field strength and duration of interaction, the relative work done on ions of charge $Z_1,Z_2$ and mass $m_1,m_2$ is $W_1/W_2=(Z_1)^2m_2/(Z_2)^2m_1$.  This suggests the typical energy of protons should be smaller than that of (fully-ionized) boron by factor of 2.5 (i.e. $T_p\simeq 0.4T_B$).  Good experimental measurements of ion kinetic energy distributions inside the laser-heated target a difficult to come by.  Fortunately we find that the yield is mostly sensitive to the absolute kinetic energy scale, controlling how much of the ion distribution is above the threshold CM energy determined by the cross section.  Once the threshold CM energy is achieved by a majority of the distribution, the yield becomes less sensitive to further increases in kinetic energy.  At next order, the yield \eq{Ythermalrate} is more sensitive to the kinetic energy of the heavier ion, due to the residual exponential dependence on $\nu$.  Note however that these yields are likely to increase somewhat for higher kinetic energy range if cross section data across a wider range of CM energy were available.  These results are exhibited in \fig{DIyieldratios}.

The relative insensitivity of $\isotope[10]{B}(p,\gamma)\isotope[11]{C}$ to ion mean kinetic energy is probably an artifact of the limited data range available for the cross section, which causes the thick target yield to plateau rapidly above $\sim 8$ MeV.  The greater sensitivity of $\isotope[11]{B}(p,\gamma)\isotope[12]{C}$ to the mean kinetic energy could allow the yield ratio to be used in a more conventional manner: by measuring both yields in the ratio, one determines to high accuracy the mean kinetic energy of ions in the target.  This method is in fact how yield ratios are commonly used in heavy-ion collisions \cite{Letessier:2002gp}.  In particular, if photons from both processes $\isotope[11]{B}(p,\gamma)\isotope[12]{C}$ and $\isotope[10]{B}(p,\gamma)\isotope[11]{C}$ can be detected, the ratio of these photon yields alone could probe the mean kinetic energy of ions in the target.  We expect the sensitivity of the photon ratio can only be established with more cross section data.

\subsection{Beam-target}

More experimental information is available on the inputs for the beam-target setup.  With a $\sim 1/2$ reduction to the total yield, experiments can measure the laser-produced ion beam on-shot.  For example, the target can cover roughly half the solid angle of the beam, so that the other half the beam propagates unperturbed into a diagnostic.  Since ion acceleration mechanisms are azimuthally symmetric or at most display a dipole azimuthal mode (for example in BOA \cite{yin2011three}), we can infer the distribution in the unmeasured half by mirroring the measured half.  Experiments on the TPW and else frequently show single or double Maxwellian ion spectra.  For simplicity and clarity, we consider a single Maxwellian distribution describing the beam, though with much smaller $\beta$ parameter than in the DI-region ion distributions.  The yield for a double Maxwellian is a suitably weighted superposition of the yield for single Maxwellians, and the effect on the yield ratios can be naturally deduced from this.

We first compare the total $\alpha$ yield to the $\isotope[11]{C}$ yield in boron and boron-nitride targets.  As shown in \fig{alphaby11Chotvcoldlosses}, for the cross sections and expected temperature of the target, beam energy losses are well-approximated by the cold limit, which we use here.  In boron targets, $\isotope[11]{B}(p,2\alpha)\isotope[4]{He}$ is the dominant source of $\alpha$ particles, while in boron-nitride targets the $\isotope[14]{N}(p,\alpha)\isotope[11]{C}$ process can contribute a similar number.  Therefore, for BN targets we show both the total $\alpha$ to $\isotope[11]{C}$ ratio and the $\isotope[11]{B}(p,2\alpha)\isotope[4]{He}$ yield relative to the (total) $\isotope[11]{C}$ yield, so that the measured $\isotope[11]{C}$ number can be both compared to the measured $\alpha$ yield (e.g. in CR-39) and used to estimate the number of $\isotope[11]{B}(p,2\alpha)\isotope[4]{He}$ reactions occuring.

\begin{figure}
\includegraphics[width=0.5\textwidth]{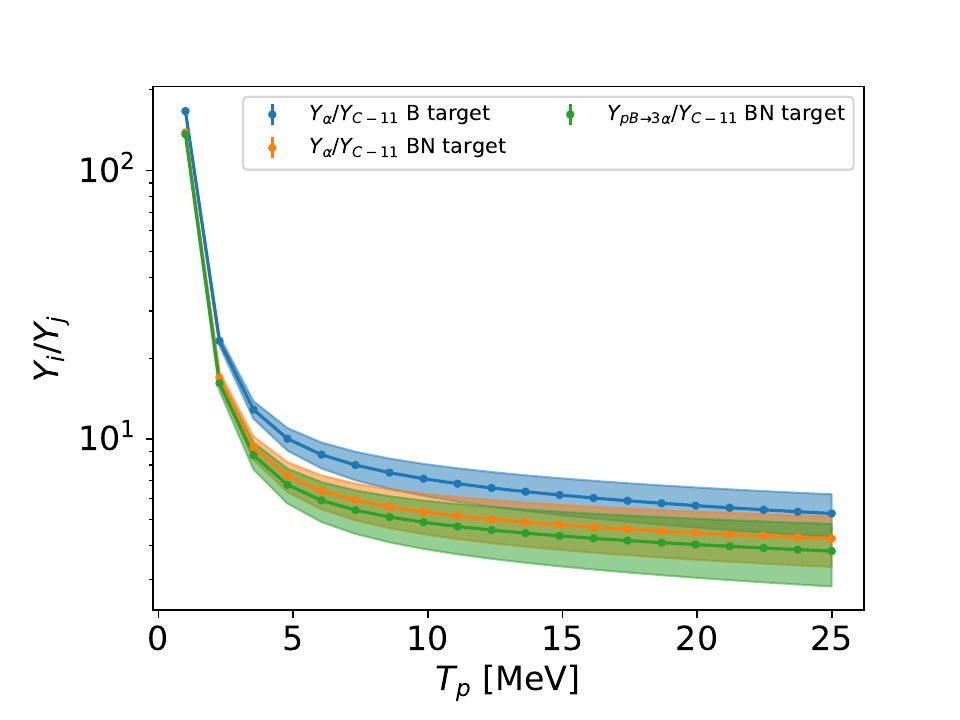}
\caption{\label{fig:BBNcomp} Yield ratios: measured total $\alpha$ and $\isotope[11]{B}(p,2\alpha)\isotope[4]{He}$ yield relative to measured $\isotope[11]{C}$ for boron and boron-nitride targets.  Beam energy losses are included with a cold target. }
\end{figure}

\begin{figure}
\includegraphics[width=0.5\textwidth]{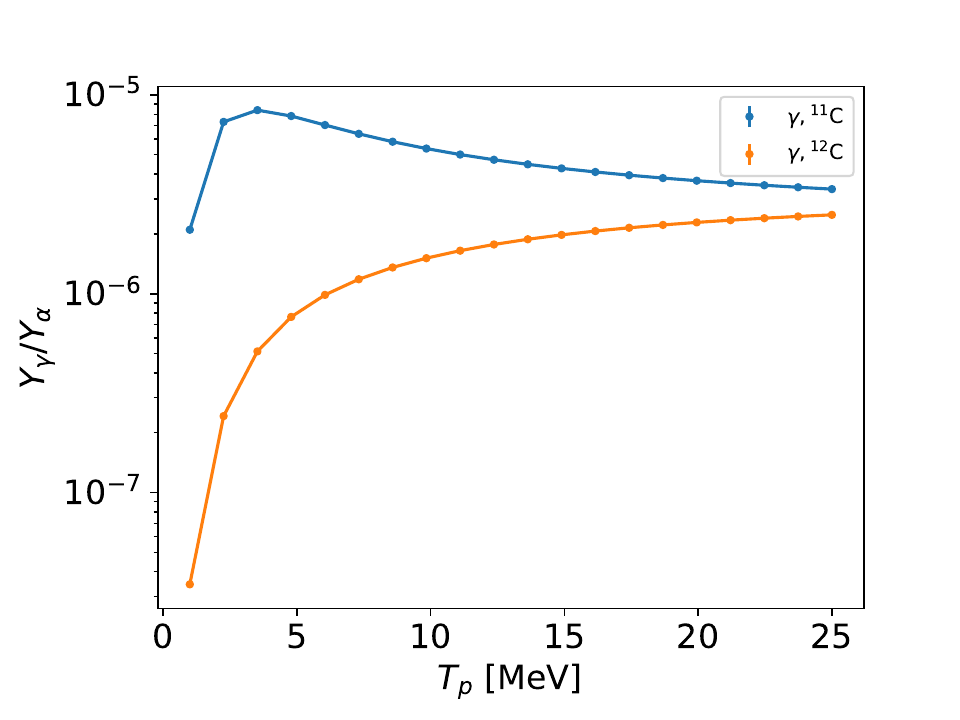}
\caption{\label{fig:gammabyalpha} Yield ratios: number of photons per $\isotope[11]{B}(p,2\alpha)\isotope[4]{He}$ reaction for the $\isotope[11]{B}(p,\gamma)\isotope[12]{C}$ and $\isotope[10]{B}(p,\gamma)\isotope[11]{C}$ processes.}
\end{figure}

Photons are also produced by the $\isotope[11]{B}(p,\gamma)\isotope[12]{C}$ and $\isotope[10]{B}(p,\gamma)\isotope[11]{C}$ processes in the beam-target geometry.  Detection of the photons in the beam-target experiment is a natural proof-of-principle/validation step before using the photon measurement to diagnose the direct-irradiation experiments.  As shown in \fig{gammabyalpha}, the photon yields are 5 to 6 orders of magnitude smaller than the $alpha$ yield, and display different dependence on the beam $\beta$ parameter in beam-target experiments compared to the quasi-thermal plasma of direct-irradiation.  The yield of $\isotope[10]{B}(p,\gamma)\isotope[11]{C}$ peaks around 3 MeV due to the narrow range of energies for which cross section data is available; this peak may disappear with more complete cross section data.

\section{Conclusion}

We have thus arrived at a list of reactions and their products with diagnostic potential.  
\begin{enumerate}
\item $\isotope[11]{C}$ is produced by three reactions in \tab{primaryreactions} with cross sections of several hundred millibarns in 5-20 MeV center-of-mass energy range.  The third channel, $\isotope[10]{B}(p,\gamma)\isotope[11]{C}$, generally contributes less than $10^{-4}$ to the total yield.   Although it requires a higher proton energy, in experiments on an ultra high-intensity laser, 1 $\isotope[11]{C}$ is produced for every 10-100 $\alpha$ particles from the $\isotope[11]{B}(p,2\alpha)\isotope[4]{He}$ reaction, see \fig{BBNcomp}.  Modulo some uncertainty in modeling the proton beam, the $\isotope[11]{C}$ yield thus offers a strong, easily-measured signal to corroborate or substitute more direct measurements of the $\alpha$ yield \cite{McCary2022}. 

\item $\isotope[11]{B}(p,\gamma)\isotope[12]{C}$ has the same initial state as the $\isotope[11]{B}(p,2\alpha)\isotope[4]{He}$, so that in both direct-irradiation and beam-target experiments, macroscopic factors such as density, volume and time, as well as physical constant prefactors cancel in the yield ratio.  The drawback to this process is its low cross section and higher threshold: only 1 photon per million $\isotope[11]{B}(p,2\alpha)\isotope[4]{He}$ reactions is expected, according to \fig{DIyieldratios}.  That suggests $\sim 10^3-10^4$ such photons were produced in recent experiments such as Refs. \cite{giuffrida2020high,bonvalet2021energetic,McCary2022}.  Pending a direct measurement though the photon should be easily distinguishable at an energy $\gtrsim 10$ MeV.  

\item $\isotope[10]{B}(p,\gamma)\isotope[11]{C}$ involves boron-10, generally leaving nontrivial prefactors in the yield ratio.  While these prefactors are expected to be order 1, due to the similar dynamics of boron-10 versus boron-11 in a laser-heated target, they introduce additional uncertainty, which also grows with the duration of the fusion.  The yield ratio \fig{DIyieldratios} predicts 1 photon per 100,000  $\isotope[11]{B}(p,2\alpha)\isotope[4]{He}$, implying roughly $10^5$ such photons in recent experiments.  The photon energy is lower, but still likely high enough in the several MeV range to be distinguishable from other plasma sources.

\item $\isotope[10]{B}(\alpha,n)\isotope[13]{N}$ has not been evaluated here, but is an excellent candidate for corroborating the $\alpha$ yield if the neutron number can be measured.  This reaction could also determine the importance of $p$ recycling by virtue of its probably near-unity yield ratio to the isospin partner reaction $\isotope[10]{B}(\alpha,p)\isotope[13]{C}$.  A rough estimate of the thick target yield for a 4 MeV $\alpha$ suggests that $\sim 10^{-5}$ of the $\alpha$s produced may be converted to neutrons by this process.  The neutron is slow enough to be easily identified by time-of-flight spectrometers.
\end{enumerate}

One other reaction could be a good proxy for $\alpha$ yield if its cross section were independently measured in conventional nuclear scattering experiments: $\isotope[14]{N}(\alpha,\gamma)\isotope[18]{F}$.  This reaction has the benefit of producing an unstable nuclide so that its yield can be first checked in beam-target experiments.  We have also noted that the ratio between the two photon-producing processes, $\isotope[11]{B}(p,\gamma)\isotope[12]{C}$ and $\isotope[10]{B}(p,\gamma)\isotope[11]{C}$, could provide a measurement sensitive to the mean kinetic energy of the ions in the plasma.  However its accuracy is currently severely limited by the little cross section data available for $\isotope[10]{B}(p,\gamma)\isotope[11]{C}$.

Considering the experimental interest and potential applications of the $\isotope[11]{B}(p,2\alpha)\isotope[4]{He}$ reaction and other laser-driven fusion reactions, we strongly recommend increasing engagement with the accelerator-nuclear physics community to improve cross section measurements and add photon and neutron diagnostics.  Particle yields and yield ratios can become a powerful tool to determine laser-driven fusion plasma conditions, in the same way they have been thoroughly developed for probing nuclear-matter plasmas.

\begin{acknowledgments}
Work performed under the auspices of the University of Texas at Austin, supported in part by the National Science Foundation under Grant No. 2108921 and the Air Force Office of Scientific Research under Grant No. FA9550-14-1-0045.  This work and a related experiment at the University of Texas, Austin were also supported in part by HB11 Energy PTY, LTD.  Experiment time at Texas Petawatt supplied by LaserNetUS.  
\end{acknowledgments}

\bibliographystyle{apsrev4-2}
\bibliography{fusion}

\end{document}